\documentclass{evn2004}
\usepackage{txfonts}
\usepackage{graphicx}
\usepackage{setspace}
\usepackage{natbib}
\bibpunct{(}{)}{;}{a}{}{,}
\usepackage{color}
\usepackage{subfigure}
\usepackage{floatflt}

\begin{document}
\title{A VLBI Study of the Gravitational Lens JVAS B0218+357}
\author{Rupal Mittal\inst{1} \and Richard Porcas\inst{1}  \and Olaf
  Wucknitz\inst{2} \and Andy Biggs\inst{3} }
\institute{ MPIfR, Auf dem H\"ugel 69, 53121 Bonn, Germany
\and University of Potsdam, Institute of Physics, Am Neuen Palais 10, D-14469
Potsdam, Germany
\and JIVE, 7990 AA Dwingeloo, Netherlands
}

\date{Received/Accepted}

\abstract
{
We present the results of phase-referenced VLBA+Effelsberg observations at five
frequencies of the gravitational lens B0218+357 to establish the precise
registration of the A and B lensed image positions.
}

\maketitle

\section{Introduction}
The potential to determine the Hubble constant, $H_{0}$,  using the double
image lens B0218+357 was established shortly after its discovery (Patnaik et
al. 1993) to be very high. This is because of the accurately measured value
for the time delay between the images, \mbox{(10.5 $\pm$ 0.4) d} (Biggs et
al. 1999) and the wealth of data coming from numerous radio and optical
observations of this source at various frequencies and epochs that provide
constraints for the lens model. Rightly so, at times it is  described  as the
\textit{`Golden Lens'}.

Yet this system presents a few `glitches'. One of them is the
steady and systematic decline in the radio image flux density ratio with
decreasing frequency.  One of the possible explanations is a
frequency-dependent source structure (the background source is conjectured to
be a blazar), combined with the magnification ratio which changes
significantly over the extent of the structure. Such changing magnification is
perhaps likely, given that the system has the smallest image separation of
$\sim$ 330 mas amongst the known galactic lenses. In the model derived by
Wucknitz (2002) using LENSCLEAN, a shift of $\sim$ 15 mas in the position of a
point-source image can produce a change in relative magnification from 4 to
2.5 (Fig. 1). Furthermore, it is indeed common for the radio spectra of AGN
jets to steepen with distance from the nucleus, and for the position of the
radio peak at the jet base to change with frequency -- the ``core
shift''. Although such a core shift should, in general, show up as a change
with frequency of the separation between the two different core images, this
effect is insensitive to core shifts in some directions. An unambiguous
registration of the VLBI structures of the radio images at different
frequencies can show whether this effect is present, and can only be made
using the technique of phase referencing.     

\begin{figure}[h]
\resizebox{\hsize}{!}{\includegraphics{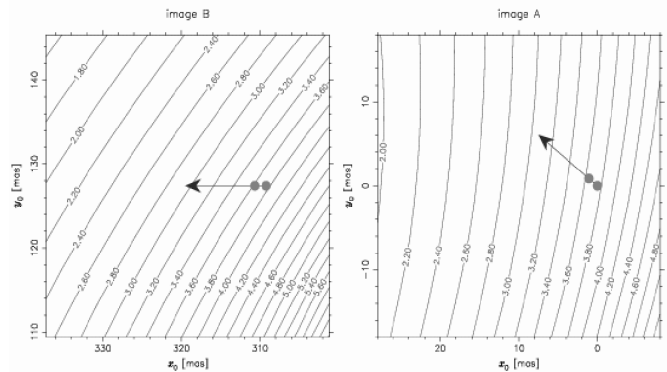}}
\caption{\small The curves indicate constant relative magnifications (A/B)
    for the best fitting lens model with the lens position at \mbox{$x_o =
  260$ mas} and $y_o = 117.5$ mas (Wucknitz, 2002)}
\end{figure}

\section{Observations}
The \textit{phase-referencing} technique is used to correct
interferometer phase errors (geometrical, atmospheric or ionospheric and
instrumental). These errors are first determined by observation of a strong, 
near-by \textit{phase (position) reference} calibrator source, which is
preferrably point-like with a frequency independent position and then
interpolated to times at which the  (usually weaker) \textit{target source} is
observed, in order to determine its structure and position relative to the
reference. In this way the derived relative geometrical offset between the
lens and the position reference is a constant only if the respective brightest
points maintain their positions with varying frequencies.  

In the case of B0218+357, the target source (the lens) is sufficiently strong
($\sim$ 1 Jy) to invert its role as the phase-reference, thereby leading to
``Inverse Phase Referencing". 

The observations were taken on the 13th and 14th of Jan. 2002 using the
VLBA (Very Long Baseline Array) and Effelsberg (Eb) at five 
frequencies, namely \mbox{15.35 GHz}, \mbox{8.40 GHz}, 4.96 GHz, 2.25 GHz and
1.65 GHz. Apart from observing the lens, three position-reference sources were
observed along with a fringe finder. The data were correlated at the VLBA
correlator and further processed in AIPS.

\begin{figure*}[t]
\centering
\includegraphics[width=17cm,height=4.8cm]{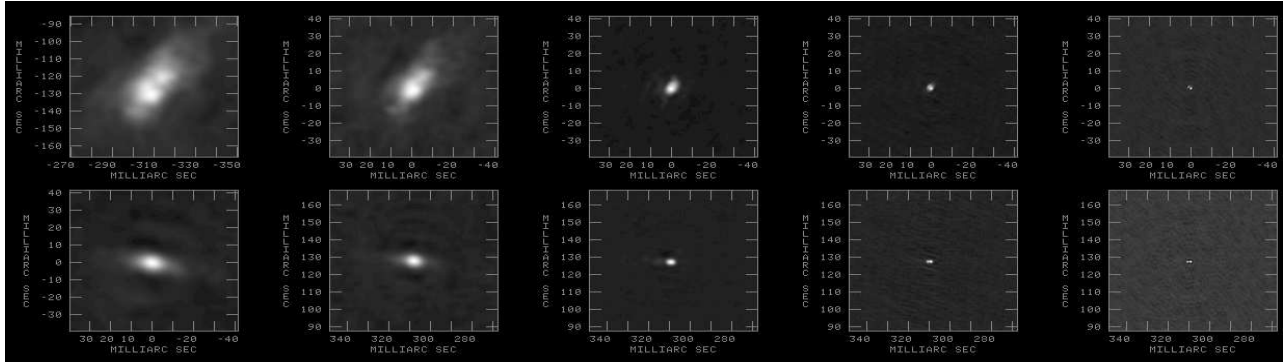}
\hfill        
\caption{\small{ Image A (top) and image B (bottom) at 1.65 GHZ, 2.25 GHz,
    4.96 GHz, 8.40 GHz, 15.35 GHz (from left to right) plotted with
  the same beam size with each side of the square measuring 80 mas. } }
\end{figure*}

\section{Maps of B0218+357}
\textit{Hybrid maps} (using phase self-calibration techniques) of the
lens were made by cleaning the two sub-fields containing the two images A and
B (separated by $\sim$ 334 mas) simultaneously. The images clearly manifest
all the earlier observed  lensing-characteristics, such as image A
being tangentially stretched at a PA $\sim$ $-40\,^{\circ}$
(Fig. 2). At 8.4 GHz and higher
frequencies, the images are resolved further into two sub-components,
separated by about 1.4 mas, representing the core-jet morphology of the
background source.

\section{Phase Referencing}
For the phase-reference analysis only one source, 0215+364, was chosen as the
most appropriate \textit{position reference}  based on its flat
spectrum in comparison with the others and also being in a suitable enough
brightness range for the purpose of determination of the brightest component
unambiguously. The hybrid maps of the images A and B were used to
investigate the change in their positions with respect to 0215+364 as a
function of frequency. Fig. 3 indicates a shift of only $\le 2$
 mas in the peaks of the image radio emission between 15.35 GHz and \mbox{1.65
  GHz}, comparable to the separation between the two sub-components seen in
both images at 15.35 GHz. Over this distance the change in relative
magnification is expected to be small.    

\begin{figure}
\centering
\includegraphics[width=6.5cm,height=5.8cm]{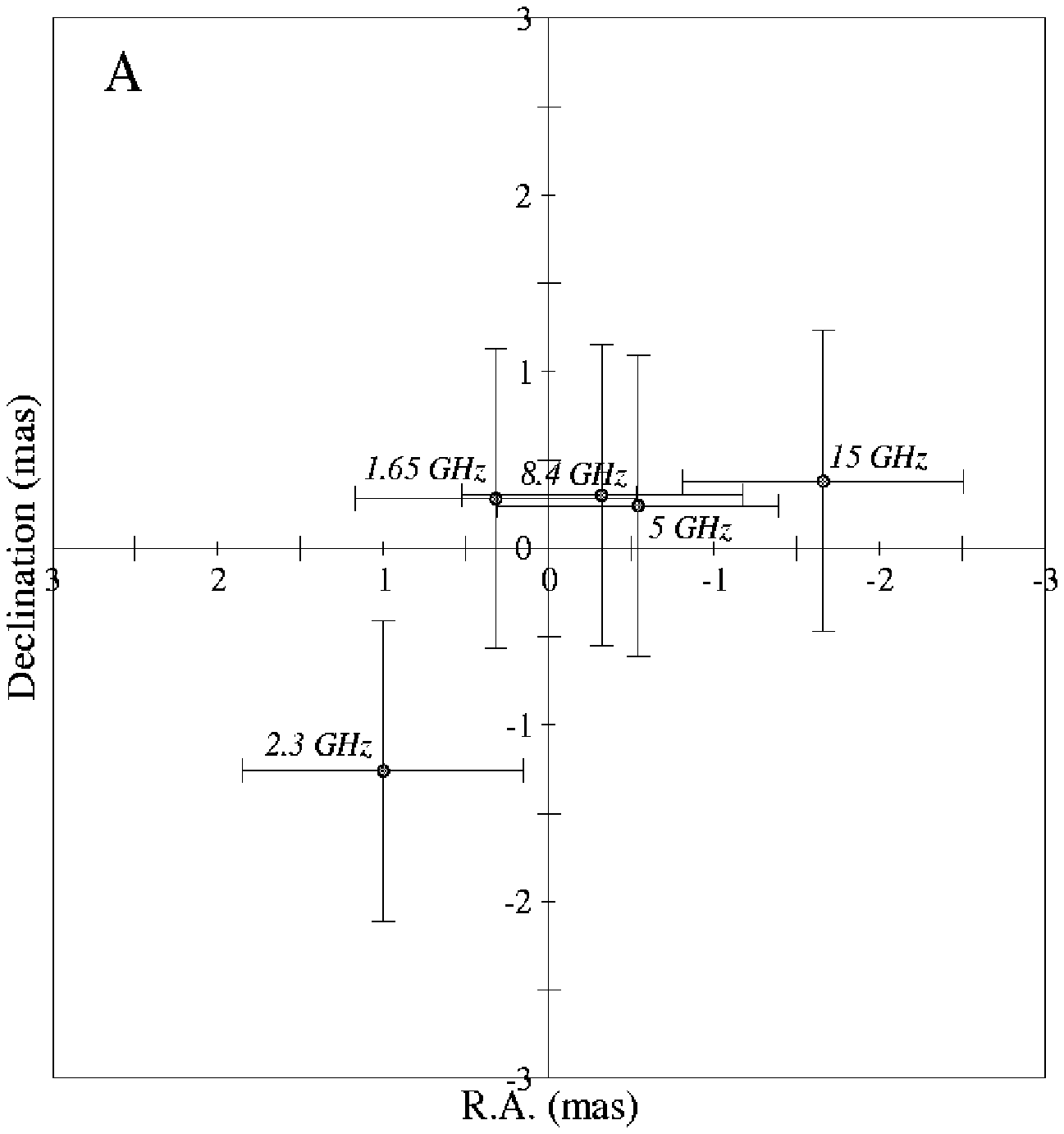}  
\includegraphics[width=6.5cm,height=5.8cm]{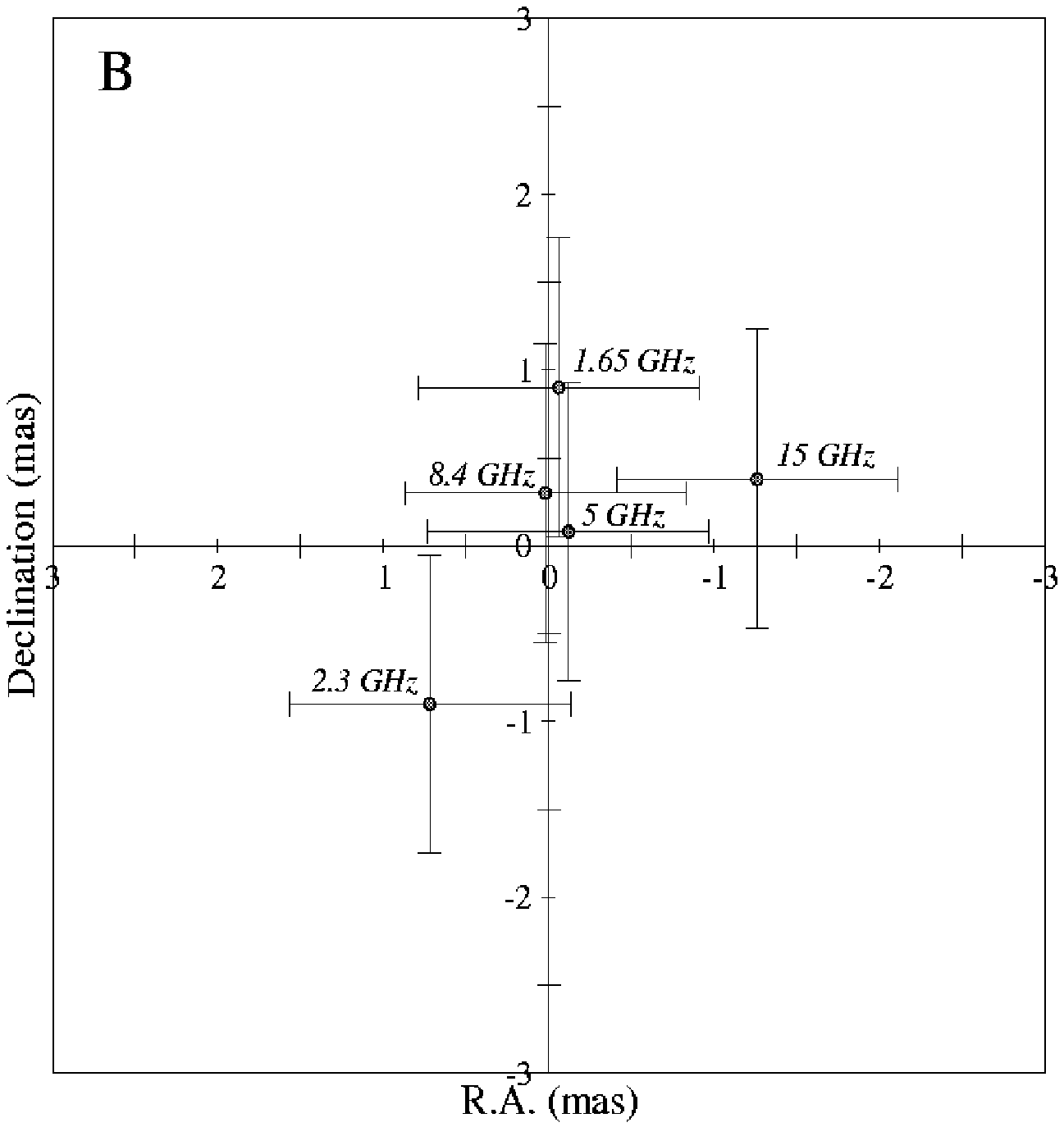}
\caption{\small The top panel shows the change in position of the peak in
image A,  relative to the position-reference 0215+364, with frequency. The
bottom panel shows the same for image B. }
\end{figure}

\section{Discussion}
Since there is no measured shift with frequency of either image peak
positions large enough to account for the anomalous flux density ratios, this
effect may be due to the frequency-dependant source-size as seen in the two
images. The different image sizes at varying frequencies could also
result from scattering in the lens galaxy (see Biggs et al. 2002 for further
discussion). At 1.65 GHz there is a relatively huge amount of low brightness
emission that extends out to $\sim$ 30 mas, in comparison to \mbox{15.35 GHz}
where the emission is dominated by the compact  sub-components with a
separation of $\sim$ 1.4 mas.  Since at lower  frequencies the (larger)
images extend over regions where lens models predict significant changes in
the relative magnification, the image flux densities(and their ratio) do not
result from the integral of their radio brightness over the entire structure
with a constant magnification.  We are therefore attempting to use the
optimal lens mass model derived from  LENSCLEAN (Wucknitz 2002) to calculate
the magnifications for discrete regions in the image plane. We can then
compare the ratio of the predicted averaged magnification over image A to
that of B for each of the frequencies, to the observed ratio of the image
flux densities.     

Alternatively, sub-structure in the lens galaxy might cause larger
gradients in the image magnifications than are found for a smooth lens
model (Fig.1)          

\bibliographystyle{aa}

\end{document}